\documentclass[11pt,a4paper]{article}
\usepackage{jheppub}
\usepackage{lipsum}
\usepackage{braket}
\usepackage{tensor}
\usepackage{dsfont}
\usepackage{tcolorbox}
\usepackage{caption}
\usepackage{subcaption}
\usepackage{tikz}
\usetikzlibrary{decorations.pathreplacing,calligraphy}
\usetikzlibrary{decorations.pathreplacing}
\usetikzlibrary{decorations.pathmorphing, patterns,shapes}
\usepackage{bm}
\usepackage{caption}
\usepackage{multirow}
\usepackage{subcaption}
\usepackage{ragged2e}
\usepackage{cancel}
\usepackage{comment}
\usepackage{xcolor}
\DeclareCaptionJustification{justified}{\justifying}
\usepackage{hyperref}
\usepackage{amssymb}
\usepackage{tikz}
\usetikzlibrary{decorations.pathreplacing,calligraphy}
\usetikzlibrary{decorations.pathreplacing}
\usetikzlibrary{decorations.pathmorphing, patterns,shapes}
\usepackage{bm}

\newcommand{\op}[1]{\boldsymbol{#1}}
\newcommand{\de}{\mathrm d}

\title{Ginzburg--Landau description for multicritical Yang--Lee models}

\author[a,b]{M\'at\'e Lencs\'es}
\affiliation[a]{HUN-REN Wigner RCP, Konkoly-Thege Miklós út 29-33, 1121 Budapest, Hungary}
\affiliation[b]{Department of Theoretical Physics, Institute of Physics, Budapest University of Technology and Economics, H-1111 Budapest, M{\H u}egyetem rkp. 3.}
\author[c]{Alessio Miscioscia}
\affiliation[c]{
 Deutsches Elektronen-Synchrotron DESY, Notkestr. 85, 22607 Hamburg, Germany
}%
\author[d]{Giuseppe Mussardo}
 \affiliation[d]{
 SISSA \& INFN, Sezione di Trieste, via Bonomea 265, I-34136, Trieste, Italy
}%
\author[b,e,f]{G\'abor Tak\'acs}
\affiliation[e]{MTA-BME Quantum Correlations Group (ELKH), Institute of Physics, Budapest University of Technology and Economics, H-1111 Budapest, M{\H u}egyetem rkp. 3.}
\affiliation[f]{
BME-MTA Momentum Statistical Field Theory Research Group, Institute of Physics, Budapest University of Technology and Economics, H-1111 Budapest, M{\H u}egyetem rkp. 3.}
\emailAdd{ mate.lencses@gmail.com}
\emailAdd{alessio.miscioscia@desy.de}
\emailAdd{mussardo@sissa.it}
\emailAdd{takacs.gabor@ttk.bme.hu}
\preprint{}
\abstract{
We revisit and extend Fisher's argument for a Ginzburg--Landau description of multicritical Yang--Lee models in terms of a single boson Lagrangian with potential $\varphi^2 (i \varphi)^n$. We explicitly study the cases of $n=1,2$ by a Truncated Hamiltonian Approach based on the free massive boson perturbed by  $\op P \op T$ symmetric deformations, 
providing clear evidence of the spontaneous breaking of $\op P \op T$ symmetry. 
For $n=1$, the symmetric and the broken phases are separated by the critical point corresponding to the minimal model $\mathcal M(2,5)$, while for $n=2$, they are separated by a critical manifold corresponding to the minimal model $\mathcal M(2,5)$ with $\mathcal M(2,7)$ on its boundary. Our numerical analysis strongly supports our Ginzburg--Landau descriptions for multicritical Yang--Lee models.
} 
\keywords{Field Theories in Lower Dimensions, Renormalization Group, Scale and Conformal Symmetries.}
\date{April 4, 2024}

\preprint{DESY-24-032}

\begin{document}
    \maketitle
	\flushbottom
 \section{Introduction and summary}

The study of critical phenomena is crucial for the theoretical understanding of quantum field theories and the experimental measurements \cite{Mussardo:2020rxh}. It is well known that many critical points are strongly coupled quantum field theories ruled by conformal symmetry.
In two space-time dimensions, the conformal group is enhanced to an infinite symmetry algebra, namely the Virasoro algebra \cite{Belavin:1984vu}. 

Among the irreducible representations of the Virasoro algebra, it is natural to consider the minimal models $\mathcal M(p,q)$, which give rise to an infinite class of two-dimensional conformal field theories. 
Those models are labelled by two coprime positive integers $p$ and $q$ and have central charge
\begin{equation}
    c = 1-6\frac{(p-q)^2}{pq} \,.
\end{equation}
Among the minimal models, the only theories which are unitary are those in the sequence $\mathcal M(2+n,3+n)\quad n=1,2,\dots$. In all other cases, unitarity is lost due to a real but non-positive spectrum of conformal weights which leads to states of negative norm.

Based on the concept of universality, one might expect that some features of strongly coupled conformal models can be captured through a weakly coupled description. Indeed, what matters is the symmetry of the problem: a weakly coupled theory respecting the same symmetry can also describe certain aspects of the same fixed point.
This approach is known as Ginzburg--Landau description~\cite{Landau:1937obd}, and it is well realised by the Ising model, which is the Wilson-Fisher fixed point in $d\le 4$ \cite{Wilson:1971dc}.

The idea has been further extended to unitary multicritical points in two dimensions. Indeed, 
in~\cite{Zamolodchikov:1986db} Zamolodchikov proposed the Ginzburg--Landau description of unitary minimal models with diagonal modular invariants in terms of bosonic Lagrangians:
\begin{equation}
    \label{eq:multiIsing_GL}
    \mathcal L _{\mathcal M(2+n,3+n)}= \frac{1}{2}(\partial \varphi)^2+ g \varphi^{2 (n+1)}\,.
\end{equation}
Besides the above sequence, the general Ginzburg--Landau description of minimal models is not presently known, despite a handful of exceptions, which are the following:
\begin{itemize}
    \item[$\star$] The minimal model $\mathcal M(2,5)$, also known as Yang--Lee model, can be described by the Ginzburg--Landau lagrangian \cite{Fisher:1978pf}  
\begin{equation}\label{eq:YL_GL}
    \mathcal L _{\mathcal M(2,5)}= \frac{1}{2}(\partial \varphi)^2+ i g \varphi^{3}\,,
\end{equation}
    \item[$\star$] The minimal model $\mathcal M(3,8)$, also known as the supersymmetric version of the Yang--Lee model, can be described by the Ginzburg--Landau lagrangian \cite{Klebanov:2022syt}
\begin{equation}
    \mathcal L _{\mathcal M(3,8)}= \frac{1}{2}(\partial \varphi)^2+ \frac{1}{2}(\partial \sigma)^2+ i g_1 \sigma \varphi^{2}+i g_2 \sigma^3\,,
\end{equation}
in terms of the two bosonic fields $\varphi(x)$ and $\sigma(x)$.
	\item[$\star$] A suggestion for the Ginzburg--Landau description of the minimal model $\mathcal M(3,10)$ was also given in \cite{Kausch:1996vq} and \cite{Klebanov:2022syt} as two copies of~\eqref{eq:YL_GL} i.e. the Yang--Lee theory.
\end{itemize}

Recently we proposed that the sequence of the non-unitary minimal models $\mathcal M(2,2n+3)$ generalizes the Yang--Lee edge singularity to its multicritical versions~\cite{Lencses:2022ira,Lencses:2023evr}.
Moreover, we suggested the existence, and provided numerical evidence in several cases, of the following RG flows from unitary multicritical points to the non-unitary Yang--Lee multicritical CFTs as well as flows between non-unitary the multicritical points:
\begin{align}
		\mathcal M(2+n,3+n) &\to \mathcal M(2,2m+3) \hspace{1 cm} m\le n \,,\label{eq:U_NU}\\
	\mathcal M(2,2n+3) &\to \mathcal M(2,2m+3) \,,\hspace{1 cm}m < n \,,\label{eq:NU_NU}
\end{align}

This paper focuses on the Ginzburg--Landau description of the series of non-unitary minimal models $\mathcal M(2,2n+3)$. The central charge $c$ and the effective central $c_{\text{eff}}$ of these models are 
\begin{equation}
    c = 1- 3\frac{(1+2n)^2}{2n+3} \,,\hspace{1 cm} c_\text{eff}= c-24 \Delta_\text{min} = \frac{2n}{2n+3} \,,
\end{equation}
The $n+1$ Virasoro primary fields $\phi_k$, $k=1,\dots,n+1$ have conformal weights
\begin{equation}\label{eq:confWieght}
    \Delta_{\phi_k}= \frac{(2n+3-2k)^2-(2n+1)^2}{8 (2n+3)} \le 0 \,,
\end{equation}
$k=1$ corresponds to the identity, while the other $n$ fields are the nontrivial relevant fields of these classes of universality. Motivated by the a combination of the above spectrum of conformal fields, Fisher's argument to the Yang--Lee model and Zamolodchikov's proposal for the unitary case, in this paper we propose that the Ginzburg--Landau Lagrangian for the universality classes of this series is given by 
\begin{equation}\label{eq:proposal}
    \boxed{\mathcal L_{\mathcal M(2,2n+3)} = \frac{1}{2}(\partial \varphi)^2+ i g_1 \varphi
    +\sum\limits_{k=0}^{n-2} g_{k+2}\varphi^2 \left(i \varphi\right)^k
    +\varphi^2 \left(i \varphi\right)^n} \,.
\end{equation}
This model has $n+1$ relevant fields, given by the powers of the elementary scalar field $\varphi,\varphi^2,\dots ,\varphi^{n+1}$. The equation of motion makes one of these fields redundant, leading to the same counting as in $\mathcal M(2,2n+3)$. As we show in the sequel, the proposed Lagrangian \eqref{eq:proposal} identifies the class of universality, whose $n$th multicritical point can be reached by tuning the couplings $g_k$ of the subleading powers. The full phase diagram also contains submanifolds of lower multicriticality, and the $k+1$th multicritical points are conjectured to form the boundary of the manifold of $k$th multicritical points.

The flows~\eqref{eq:U_NU} and~\eqref{eq:NU_NU}, when expressed in terms of the highest power in the field potential, schematically correspond to the flows 
\begin{align}
	\varphi^{3n} &\to \varphi^2 (i\varphi)^m \,,\\
	 \varphi^2 (i\varphi)^n  &\to   \varphi^2 (i\varphi)^m  \,.
\end{align}
We provide arguments supporting this proposal and numerically check it for the first two cases: the Yang--Lee fixed point and its tricritical version.
It is important to notice that the above Lagrangians are explicit $\op P \op T$ symmetric, and this important feature is the key point which guarantees the reality of the conformal spectra of the associated fixed points. In addition, the Lagrangians in \eqref{eq:proposal} are the field theory generalization of the quantum mechanical models initially studied in \cite{Bender:1998ke}, which were the first examples of non-Hermitian, but $\op P \op T$ symmetric model with real spectrum. In this perspective, our proposal is a field-theoretic generalization of those theories. Even if the spectrum of $\op P \op T$ symmetric models is real, there can be negative norm states. In the quantum mechanical setting, the negativity of the norm of those states is cured by defining a new inner product (called the $\op C \op P \op T$ inner product, see discussion in Section B of \cite{Bender:2023cem} and references therein). Defining the equivalent of the $\op C$ operator in this field theory framework is a very interesting problem, which is not the goal of this paper; nevertheless, we make some short comments in this regard. In the CFT context, the natural inner product is defined by the Virasoro algebra \cite{Belavin:1984vu,Mussardo:2020rxh}, and the negativity of the norm is treated as an expected feature of the models. 
The inner product defined in the CFT naturally extends to the off-critical theory. Consequently, the residues of the poles of the S-matrix are not positive, and the completeness relation has to be dressed with negative signs. This is easily demonstrated using the scaling region of the Yang-Lee model, where negative signs must be considered in front of the form factor contributions corresponding to an odd number of particles \cite{1989PhLB..225..275C,Zamolodchikov:1990bk}. Hence, concerning the correlation functions, it seems more important to define the completeness relation rather than pursuing a definition of a positive scalar product.

The paper is structured as follows: in Section \ref{sec:proposal}, we revisit and extend Fisher's argument \cite{Fisher:1978pf}, in favour of our proposal~\eqref{eq:proposal}, then we elaborate on the role of $\op P \op T$ symmetry.
In Section \ref{sec:Hamilt} we set up the Hamiltonian truncation for the Ginzburg--Landau models, and test its correct implementation   
using known results regarding the massive free boson and the Wilson--Fisher fixed point. We locate the latter fixed point in the Chang dual channel and compared the ratio between critical couplings with the expected value, finding good agreement with the expected results.
We then study the scaling region of the first two cases, which correspond to the leading interaction terms $i g \varphi^3$  and $g \varphi^2 (i \varphi)^2$, and we numerically confirm the validity of our proposal involving the Ginzburg-Landau lagrangians presented in equation \eqref{eq:proposal} for $n = 1$ and $n = 2$. We discuss the results and our conclusions in Section \ref{sec:discussion}.

\section{Proposal and arguments}\label{sec:proposal}

Our proposal~\eqref{eq:proposal} is the natural generalization of the well-established Yang--Lee result~\eqref{eq:YL_GL} \cite{Fisher:1978pf,Cardy:1985yy,Xu:2023nke}: notice that it contains the necessary number of relevant degrees of freedom, it respects $\op P \op T$ symmetry and is compatible with the RG flows found in~\cite{Lencses:2022ira,Lencses:2023evr}.

In this section, we revisit and extend Fisher's argument to support our proposal. Then we present further arguments in line with the expectations from the bootstrap description of integrable massive deformations. Finally, we elaborate on the role of $\op P \op T$ symmetry and the general expectation for the phase diagram related to its spontaneous breaking.

\subsection{Fisher's argument revisited}\label{subsec:Fisher}
The Ginzburg--Landau description of the Yang--Lee edge singularity is given by
\begin{equation}\label{eq:LeeYangGL}
	\mathcal L_\text{Yang--Lee}= \frac{1}{2}(\partial \varphi)^2+ i g_1 \varphi + \varphi ^2 (i \varphi)  \,.
\end{equation}
Fisher obtained this result in \cite{Fisher:1978pf}, which was later used by Cardy to argue that the Yang--Lee edge singularity corresponds to the conformal minimal model $\mathcal M(2,5)$ \cite{Cardy:1985yy}. 
Here we review and generalise Fisher's method.
This was already attempted in \cite{vonGehlen:1994rp}, but our recent results obtained in \cite{Lencses:2022ira,Lencses:2023evr} on the non-unitary models guide us to get the correct generalisation.

Fisher's idea is to start from the Ginzburg--Landau description of Ising, which is just a $\varphi^4$ theory
\begin{equation}
    \mathcal L_{\mathcal M(3,4)} = \frac{1}{2}(\partial \varphi)^2+ g_1 \varphi+g_2 \varphi^2 + \varphi^4 \,.
\end{equation}
The $\varphi^3$ term is absent in this Lagrangian due to an appropriate field shift.
Following the argument of Zamolodchikov \cite{Zamolodchikov:1986db} based on the OPE structure of the Ising fixed point, it is clear that $\varphi$ is mapped to the magnetic field $\sigma$ of the minimal model $\mathcal M(3,4)$, while $\varphi^2$ is mapped to the energy operator $\epsilon$. 
This is further supported by a simple symmetry argument: $\sigma$ is a $\mathbb Z_2$ odd field while $\epsilon$ is $\mathbb Z_2$ even.

In \cite{Yang:1952be,Lee:1952ig} Lee and Yang showed that a new critical point, the Yang--Lee edge singularity, arises when the Ising model is deformed with an imaginary magnetic field setting $g_1$ to $i g_1$. As Fisher showed in \cite{Fisher:1978pf}, this can be implemented in the Ginzburg--Landau description by the following steps: 
\begin{enumerate}
\item shifting the field as 
\begin{equation}
    \varphi\to \varphi+ i \varphi_0 \,,
    \label{shiftfielf}
\end{equation}
\item setting $g_2 \to 6 \varphi_0^2$ in order to set the $\varphi^2$ term equal to zero; 
\item neglecting the constant terms and the $\varphi^4$ term since this operator becomes irrelevant in the infrared theory.  In this way, we end up with the Lagrangian in \eqref{eq:LeeYangGL}.
\item In particular, the theory of the new critical point is defined by tuning the coupling in front of $\varphi$ to set the mass gap to zero. This can be achieved by setting $i g_1 \to -8 \varphi_0$. The resulting Lagrangian is a $\varphi^3$ theory with an imaginary coupling (equal to $4 i \varphi_0$ ).
\end{enumerate}

Now we demonstrate that the same procedure can also be performed starting from the GL Lagrangian of the tricritical Ising model: 
\begin{equation}
    \mathcal L_{\mathcal M(5,6)} = \frac{1}{2}(\partial \varphi)^2+ g_1 \varphi + g_2 \varphi^2+g_3 \varphi^3+g_4 \varphi^4+\varphi^6 \,.
\end{equation}
We should follow steps similar to the Yang-Lee case: \begin{enumerate}
    \item shifting the field as 
\begin{equation}
    \varphi\to \varphi+ i \varphi_0 \,.
    \label{shiftfielf}
\end{equation}
we arrive at the following Lagrangian:
\begin{equation}
     \mathcal L = \frac{1}{2}(\partial \varphi)^2+ i \gamma_1 \varphi + \gamma_2 \varphi^2+i  \gamma_3 \varphi^3+\gamma_4 \varphi^4+i\gamma_5 \varphi^5+\varphi^6 \,.
\end{equation}
where the $\gamma_k$ depend on the original couplings $g_n$, when we consider imaginary couplings in front of the odd powers of $\varphi$, and the shift of the field $\varphi_0$.
\item setting $g_1 \to -6 \varphi_0^5-2\varphi_0 g_2+3 \varphi_0^2 g_3+4 \varphi_0^3 g_4$ we can implement the condition $\gamma_1 = 0$.
\item setting $g_2 \to -3\left(5 \varphi_0^4-\varphi_0 g_3-2 \varphi_0^2 g_4\right)$ we can enforce the condition $\gamma_2 = 0$, reach in this way the first universality class. This corresponds to the universality class of the Yang--Lee edge singularity universality since, once we neglect the constant terms and the higher powers of $\varphi$, the Lagrangian has a $\varphi^3$ potential with an imaginary coupling in front (specifically $i (-20 \varphi_0^3+g_3+4 \varphi_0 g_4)$). So, the first conclusion is that there is a fixed point of the Yang--Lee type in the tricritical Ising model as well; as a matter of fact, these fixed points form a line.
\item Since the tricritical Ising model has more relevant degrees of freedom than the Ising model, this circumstance allows us to tune one more coupling. We can adjust $g_3 \to 4 \left(5 \varphi_0^3-\varphi_0 g_4\right)$ to set the $\varphi^3$ term equal to zero and reach in this way the second universality class. The resulting Ginzburg--Landau Lagrangian takes the form 
\begin{equation}\label{eq:phi^4}
    \mathcal L = \frac{1}{2}(\partial \varphi)^2+\ldots+ \gamma_4 \varphi^4\,,
\end{equation}
where $\gamma_4 = -15 \varphi_0^2+g_4$.
\end{enumerate}
The naive expectation is that the latter Langragian should describe the tricritical version of the Yang--Lee fixed point at the end of the line of ordinary Yang--Lee fixed points.
It was shown in \cite{Lencses:2022ira} that in the $\op P \op T$ symmetric sector of the scaling region of the tricritical Ising, a manifold of conformal points of the Yang--Lee type ends in a fixed point of a different universality class, which is a tricritical point proposed to be the tricritical version of the Yang--Lee edge singularity. The Lagrangian in \eqref{eq:phi^4} seems naively unitary and to coincide with the Ginzburg--Landau theory of Ising, at least when $\gamma_4>0$. 
However, it is known that there is a $\op P \op T$ invariant version of this theory which is obtained for $\gamma_4<0$. Although at first sight this seems to correspond to an unstable potential, if one adopts a different quantization condition corresponding to an analytical continuation to complex contours in the $\varphi$ plane, the Lagrangian which was written above can be considered as the $n =2$ case of \eqref{eq:proposal}, as shown to be the case for its $0+1$-dimensional (quantum mechanical) counterpart \cite{Bender:1998ke}. We return to this line of thought in more detail in Section \ref{sec:pt}. 

One might think that implementing one more tuning of the parameters will lead us to another critical point whose Ginzburg--Landau potential is governed by $\varphi^5$ with an imaginary coupling in front.
However this is not the case, as it was argued in \cite{Lencses:2022ira}; indeed, tuning more parameters to get other critical points is impossible if the ultraviolet fixed point is the tricritical Ising, i.e. the minimal model $\mathcal M(2,9)$ according to \cite{Lencses:2022ira,Lencses:2023evr}.
The argument is as follows: to reach a critical surface (which in this case is expected to be a line of ordinary critical points) from the tricritical Ising, it is necessary to tune the mass gap to zero, i.e. to make the ground state and the first excited state meet.
The ends of this critical line are then expected to correspond to a different universality class corresponding to the non-unitary tricritical point, where the ground state meets simultaneously with both the first and the second excited states.
It is then clear that to stay on the (one-dimensional) critical line, the field shift $\varphi_0$, and the couplings $g_1$, $g_2$, $g_3$ and $g_4$ cannot be varied independently, which prevents the tuning necessary to obtain a critical point governed by an $i \varphi^5$ Lagrangian. The latter can only be obtained by starting from a Lagrangian with more parameters that can be tuned. This the case if we start from the Lagrangian corresponding to the tetracritical Ising, i.e. with the higher power given by the monomial $\varphi^8$: in this case, by shifting the field $\varphi$ as before, it is possible to tune $\varphi_0$ and the couplings in front of $\varphi$ and $\varphi^2$ to reach the first critical point, corresponding to the Ginzburg--Landau form of the Yang--Lee singularity with the highest relevant power $i\varphi^3$. Tuning the coupling in front of $\varphi^3$ term, one can then reach the Ginzburg--Landau theory with the highest term $\varphi^2 (i \varphi)^2$, corresponding to the tricritical version of Yang--Lee, expected to be $\mathcal M(2,7)$. Notice that, in this case, there are enough free parameters that can be tuned to reach the Ginzburg--Landau theory governed by $i \varphi^5$, which is expected to be the Ginzburg--Landau description of the tetracritical Yang--Lee model. 

The above argument can be straightforwardly generalised to all higher multicritical Yang--Lee CFTs $\mathcal M(2,2n+3)$.

\subsection{Argument from integrable massive deformations}

A heuristic argument that supports the proposed Ginzburg--Landau description arises by considering the integrable deformations of the minimal model $\mathcal M(2,2n+3)$. Firstly, the primary field $\phi_{k}$ satisfy the fusion rules
\begin{equation}
    \phi_{2}\times \phi_{k}=\phi_{k-1}+\phi_{k+1}
\end{equation}
with $\phi_{n+2}$ identified with $\phi_{n+1}$. This suggests that $\phi_2$ corresponds to (the renormalised version of) the elementary GL field $\varphi$, while the $\phi_{k+1}$ to $\varphi^k$. It makes full sense that $\phi_{n+2}$ is not an independent field since the power $\varphi^{n+1}$ can be expressed in terms of the lower powers from the renormalised equation of motion. 
Secondly, the above identification becomes even more plausible by noting that the naive scaling dimensions of all powers $\varphi^k$ is zero, and the exact dimensions \eqref{eq:confWieght} originate purely as anomalous dimensions from the renormalisation of the corresponding quantum field theory. Compared to the GL description of unitary minimal models \cite{Zamolodchikov:1986db}, it is natural to assume that the size of renormalisation corrections grows with the exponent $k$. The only difference of this case, compared with to the unitary series, is that for the models $\mathcal M(2,2n+3)$, all these scaling dimensions are negative, reflecting the model's non-unitarity.

Perturbing by $\phi_3$ introduces a mass term $\varphi^2$ and leads to a massive integrable field theory, with an $S$-matrix exactly determined by bootstrap \cite{1989PhLB..229..243F}. Importantly, the spectrum supports exactly $n$ particles $A_k$, which can be obtained as the bound states of the fundamental particle $A_1$. This fundamental particle can be associated with the elementary excitation of the fundamental field $\varphi$, and fusing it $n+1$ times leads back to itself, which is consistent with the presence of the defining $\varphi^2(i\varphi)^{n}$ term in the Lagrangian \eqref{eq:proposal}. The form factor bootstrap built upon this $S$-matrix is also fully consistent with the spectrum of primary fields $\phi_k$ \cite{1994NuPhB.428..655K}.

\subsection{The role of $PT$ symmetry}\label{sec:pt}

Crucially, the GL Lagrangians written above, as well as the minimal models $\mathcal M(2,2n+3)$ and their perturbations studied in \cite{Lencses:2022ira,Lencses:2023evr} are all explicitly $\op P \op T$ symmetric, i.e. the associated Hamiltonians satisfy
\begin{equation}\label{eq:[H,PT]=0}
    [\op H,\op P \op T] = 0  \,.
\end{equation}
In the context of Ginzburg--Landau theory,  the $\op P \op T$ transformations are the following:
\begin{equation}
    x \to - x \,, \hspace{1 cm} i \to -i \hspace{1 cm} \varphi \to - \varphi  \,,
\end{equation}
implying that Lagrangians of the form
\begin{equation}
    \mathcal L = \frac{1}{2}\left(\partial \varphi\right)^2+ \varphi^2\left(i \varphi\right)^n \,,
\end{equation}
are invariant under $\op P \op T$-transformations for all $n \in \mathbb R$. In fact, the Lagrangian written above are the field-theory generalisations of the well-known $\op P \op T$-symmetric quantum mechanical systems \begin{equation}\label{eq:BenderHam}
    \op H_\text{QM} = p^2 + x^2 (i x)^\epsilon \hspace{1 cm} \epsilon\in \mathbb R \,,
\end{equation}
proposed by Bender and Boettcher \cite{Bender:1998ke}.
For the quantum mechanical case it was shown that for $\epsilon>0$, the spectrum of the Hamiltonian in \eqref{eq:BenderHam} is real because of $\op P\op T$ symmetry. Arguments and rigorous proofs for the reality of the spectrum of the theory defined by the Hamiltonian in \eqref{eq:BenderHam} (when $\epsilon>0$) are given in \cite{Bender:2006wt,Dorey:2001hi,Dorey:2001uw}. This finding demonstrates that by relaxing the assumption of Hermiticity, one can still end up in quantum field theories potentially interesting from a physical point of view. 

A notable feature of the $\op P \op T$ symmetric quantum mechanical systems is that it cannot be quantized by requiring the wavefunction to vanish as $|x|\to \infty$, in contrast to the Hermitian case. This observation has important consequences for the field theory extension.
In fact, such a condition is only adequate for the regime $1 \le \epsilon < 2$ (including, therefore, the Yang--Lee case), but not for $\epsilon\ge 2$. To obtain a real spectrum, it is necessary to replace the real $x$-axis in a contour in the complex plane \cite{Bender:1998ke}.
For this reason, the theories with potentials $g x^4$ and $g x^2 (i x)^2$ (and therefore their field theory counterparts) are intrinsically different. 
For a deeper understanding, we refer to \cite{Bender:1998ke,Bender:2020gbh,Felski:2021evi} and references therein. 

Recently, there has been renewed interest in extending the results of $\op P\op T$-symmetric quantum mechanics to field theories with $\op P\op T$-invariant interaction terms, especially those of the form $\varphi^2(i\varphi)^n$ in various space-time dimensions, using diverse approaches such as perturbation theory, expansion in the exponent $n$ and functional renormalisation group methods \cite{Felski:2021evi,Bender:2021fxa,Ai:2022csx,Naon:2022xvl,Ai:2022olh,Croney:2023gwy}.

Intriguingly, $\op P \op T$-symmetry can be spontaneously broken; in fact, the reality of the spectrum is only guaranteed when \begin{equation}\label{eq:PTsymmetryonEigen}
    \op P \op T \ket{\Psi} = e^{i \alpha} \ket{\Psi} \,, \hspace{1 cm}\alpha \in \mathbb R \,,
\end{equation}
where $\ket{\Psi}$ is an eigenvector of the Hamiltonian.
Observe that the condition in equation \eqref{eq:[H,PT]=0} does not imply \eqref{eq:PTsymmetryonEigen}: if both the conditions hold, the theory is in an $\op P \op T$-symmetric phase and the spectrum is real.
When the Hamiltonian commutes with the $\op P\op T$ operators but its eigenstates are not eigenstates of the $\op P \op T$-operators, then the theory is in a spontaneously broken $\op P\op T$-phase.
In the latter phase, the spectrum is generally complex, containing complex conjugate pairs of energy levels.
In the quantum mechanical case $\op P \op T$-breaking happens when $\epsilon$ becomes negative, and the two phases are separated by the Hermitian harmonic oscillator $\epsilon =0$ \cite{Bender:1998ke}. 

Recently, $\op P \op T$ breaking has attracted attention in the framework of two-dimensional quantum field theories. In \cite{Lencses:2022ira}, it was proposed that the Yang--Lee edge singularity can be understood as the critical point separating the $\op P \op T$ symmetric phase from the spontaneously broken $\op P\op T$ phase in the $\op P \op T$ symmetric deformation of the critical Ising model. This concept was extended for the tricritical Ising and beyond for the general multicritical case. Breaking of $\op P \op T$ symmetry was also discussed in scaling regions of the minimal models $\mathcal M(2,5)$ and $\mathcal M(2,7)$ \cite{Lencses:2023evr}, where evidence of non-critical $\op P \op T$ breaking was also found. Additionally, $\op P \op T$-symmetric scaling regions of the minimal models $\mathcal M(3,5)$ and $\mathcal M(3,7)$ were studied in \cite{Delouche:2023wsl}, and two-dimensional QCD also presents similar behaviours and $\op P \op T$ breaking (see discussion in 6.3.1 of \cite{Ambrosino:2023dik}).

In this paper, we propose that the two-dimensional field theories corresponding to the quantum mechanical Hamiltonians in equation \eqref{eq:BenderHam} with $n \in \mathbb N$, play the role of Ginzburg--Landau descriptions for the non-unitary multicritical models which are related to transitions between $\op P \op T$ symmetric and $\op P \op T$ breaking phases.  

\paragraph{Irreversibility of $\op P \op T$-symmetric RG flows.}
The similarities between Hermitian and $\op P \op T$ symmetric models do not stop in the reality of the spectrum.
The irreversibility of RG flows in two spacetime dimensions can be understood as the monotonicity of the $c$-function along the RG flows \cite{Zamolodchikov:1986gt}. 
This crucial result was generalized for flows with (unbroken) $\op P \op T$-symmetry \cite{Castro-Alvaredo:2017udm} by replacing the $c$-function with \textit{effective} $c$-function, $c_\text{eff}$, defined in the critical points as 
\begin{equation}
    c_\text{eff} = c-24 \Delta_\text{min} \,,
\end{equation}
where $\Delta_\text{min}$ is the lowest among the conformal dimensions of the theory\footnote{Note that the monotonic behaviour of the $c_\text{eff}$ function generalizes the $c$-function in unitary cases. In fact, unitarity implies that $\Delta_\text{min} = 0$, and therefore the $c_\text{eff}$ is identical to $c$. However, in the non-unitary case, conformal dimensions are generally negative.}. 

In conclusion, the irreversibility of two-dimensional RG flows in the $\op P \op T$ symmetric phase can be understood as the condition 
\begin{equation}
    c_\text{eff}^{UV} \ge c_\text{eff}^{IR} \,.
\end{equation}
This condition puts stringent restrictions on potential RG flows linking different fixed points \cite{Lencses:2022ira,Lencses:2023evr,Klebanov:2022syt,Delouche:2023wsl}. For the purposes of this paper, the constraints given by the monotonic behavior of the $c_\text{eff}$ function are automatically satisfied since the effective central charge of the free boson is $1$ and all the minimal models have an effective central charge of less than one. 

\paragraph{Expectations for the phase diagram.} 
We are now ready to consider the proposed GL description \eqref{eq:proposal} in light of $\op P \op T$ symmetry.  

Firstly, we expect to find a critical point in the scaling region of the $i \varphi^3$ theory corresponding to the Yang--Lee edge singularity. Since the Lagrangian is explicitly $\op P \op T$ symmetric, $\op P \op T$ symmetry can be either an actual symmetry of the spectrum or spontaneously broken. In analogy with \cite{Lencses:2022ira,Lencses:2023evr,Delouche:2023wsl}, it is natural to expect that the critical point separating the two phases corresponds to the Yang--Lee model.

In the case of the $\varphi^2 (i \varphi)^2$ theory, we expect to find a critical line of the Yang--Lee universality class, i.e. ruled by the conformal minimal model $\mathcal M(2,5)$.  This line is expected to separate between a $\op P \op T$-symmetric phase and a spontaneously broken $\op P \op T$-phase. In particular, we know that in the $i\varphi^3$ theory, there is a critical point of the Yang--Lee universality class. Therefore we expect the critical line described above to extend this Yang--Lee criticality to nonzero $\varphi^4$ coupling.

This critical line must end at a critical point corresponding to the tricritical version of Yang--Lee, i.e., the universality class $\mathcal M(2,7)$ \cite{Lencses:2022ira}. 
Finally, it is also possible that beyond the $\mathcal M(2,7)$ tricritical endpoint, the $\op P \op T$ symmetry is broken without passing through a critical point, as explained for the case of the scaling region of the minimal model $\mathcal M(2,9)$ in \cite{Lencses:2023evr}. This picture can be straightforwardly generalized to $n>2$. 

In the following Sections, we confirm the scenario described above using a numerical method, starting our analysis with the $n = 1$ case to warm up and then turning our attention to the $n = 2$ case.

\section{Hamiltonian truncation for Ginzburg--Landau theories}\label{sec:Hamilt}

\begin{figure}
\centering
    \includegraphics[width=0.25\textwidth]{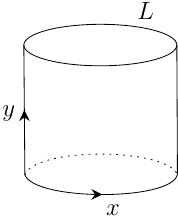}
\caption{Space-time cylinder of circumference $L$.}
\label{cylinder}
\end{figure}

To check the validity of our proposal, we implemented a non-perturbative variational approach known as the Hamiltonian truncation method. This method is particularly suited for studying  theories with discrete energy spectrum. First introduced in the context of perturbed minimal models \cite{Yurov:1989yu,Yurov:1991my,Lassig:1990xy,Lassig:1990cq}, variants of this method were then extended to more general field theories \cite{1998PhLB..430..264F,2015NuPhB.899..547K,2016JHEP...07..140K,2021JHEP...01..014K} including Ginzburg-Landau models in two \cite{2014JSMTE..12..010C,Rychkov:2014eea,Rychkov:2015vap,2016JHEP...10..050B} and higher dimensions \cite{Hogervorst:2014rta}, and situations with boundaries \cite{1998NuPhB.525..641D} and defects \cite{2009JHEP...11..057K}. Besides the computation of energy spectra, truncated Hamiltonian methods are also an efficient tool to study non-equilibrium dynamics \cite{Rakovszky:2016ugs,2019PhRvA.100a3613H,2022PhRvB.105a4305S,Szasz-Schagrin:2022wkk,Lencses:2022sfa}. 

When applying Hamiltonian truncation to the non-Hermitian Ginzburg--Landau Lagrangians \eqref{eq:proposal}, there is no a priori guarantee that it works. Fortunately, the method proves stability; more details can be found at the end of Appendix \ref{Appendix:Implementation}.

\subsection{Implementation and identification of critical points}

We start with the Hamiltonian of free massive theory in a finite volume $L$ with periodic boundary conditions, i.e., on a space-time cylinder of circumference $L$ as shown in Fig. \ref{cylinder}. The massive field $\varphi$ can be expanded in momentum modes as 
\begin{equation}\label{eq:phimodes}
    \varphi(x) = \sum_{k} \frac{1}{\sqrt{2 L \omega_k}} \left(\op a_k e^{i k x}+\op a_k^\dagger e^{-i k  x}\right) \,,
\end{equation}
where $\omega_k = \sqrt{m^2+k^2}$ is the energy of a free particle of mass $m$ with momentum $k \in 2 \pi \mathbb Z/L$.

The annihilation and creation operators obey the usual commutation rules \begin{equation}
    [\op a_{k}, \op a_{k'}] = [\op a_{k}^\dagger, \op a_{k'}^\dagger] = 0  \,, \hspace{1 cm} [\op a_k,\op a_{k'}^{\dagger}] = \delta_{kk'}  \,.
\end{equation}
and generate a Fock space from the vacuum state $\ket{0}$ which satisfies as $\op a_{k}\ket{0} = 0$:
\begin{equation}
    \ket{k_1,\ldots, k_n} = \op a_{k_1}^\dagger \ldots \op a_{k_n}^\dagger \ket{0} \,.
\end{equation}
The free Hamiltonian $\op H_0$ can be written as: \begin{equation}
    \op H_0 =\sum_{k} \omega_k \op a_k^\dagger \op a_k \,,
\end{equation}
The interaction term is a polynomial of the field $\varphi$ expressed as a linear combination of the operators
\begin{equation}
    \op V_n = \int_0^L : \varphi^n : \ \de x  \,,
\end{equation} 
and for our purposes, we limit ourselves to $n\leq 4$. In Appendix \ref{sec:appendixTerms}, all these terms are explicitly written in terms of the creation/annihilation operators \eqref{eq:phimodes}.

The most general Hamiltonian we construct is therefore \begin{equation}\label{eq:Hamiltonian}
    \op H = \op H_0+ i g_1 \op V_1+ g_2 \op V_2+ i g_3 \op V_3+ g_4 \op V_4+\ldots  \,,
\end{equation}
where $g_k \in \mathbb R$, and the $\ldots$ indicate finite volume corrections discussed later. The choice of purely imaginary couplings in front of odd powers of $\varphi$ and purely real couplings in front of the even powers of $\varphi$ ensures the $\op P \op T$ symmetry of the Hamiltonian in \eqref{eq:Hamiltonian}, since $\op P \op T$ acts as $ \varphi \to - \varphi $ and $i \to -i $. 

The key idea of the truncation approach is to construct the Fock space up to a certain energy cutoff $\Lambda$ in terms of the eigenvalue of the free field Hamiltonian $H_0$. This corresponds to splitting the Hilbert space (the Fock space) into a low-energy and a high-energy sector: \begin{equation}
    \mathcal H = \mathcal  H_l \oplus \mathcal H_h \,,
\end{equation}
and only generate the states in the low-energy sector $\mathcal H_l$, defined as:
\begin{equation}
    \ket{k_1,\ldots k_n} \in \mathcal H \,, \hspace{1 cm} \sum_{i = 1}^n \omega_{k_i} \le \Lambda \,.
\end{equation}
In particular, this automatically implies that we only have to consider only a finite number of momentum modes which satisfy  
\begin{equation}
    \frac{4 \pi^2 n^2}{L^2} = k_n^2 \le \Lambda^2-m^2  \,, 
\label{eq:cutoff_condition}\end{equation}
and so we can identify a maximum $k_\text{max}$ and/or a maximum momentum quantum number $n_\text{max}$. Since we perform our calculations in the zero-momentum subspace, the value of $k_\text{max}$ resp. $n_\text{max}$ can be taken as half the value required by \eqref{eq:cutoff_condition}, since the lowest energy level with zero total momentum and non-zero occupation number for momentum $k_\text{max}$ corresponds to the state 
$\ket{-k_\text{max},k_\text{max}}$. 

In the following calculations, we define our Hilbert space by fixing $n_\text{max}$, which has the advantage that the dimensionality of the space does not grow with the volume $L$. However, note that this implies that the energy cutoff $\Lambda$ decreases when the volume increases. As we explain later, to locate critical points and determine the conformal dimensions of primary fields, we need the large-$L$ behaviour of the lowest-lying energy levels. Therefore, the idea is to maximize precision via the number of states retained after truncation in the case of large volumes. We then choose our cutoff for lower volumes so that the dimension of the truncated space is approximately the same for all volumes.
In our following calculations, we use $n_\text{max}= 10$, corresponding to retaining thousands of states when $m L = 10$. The matrix elements of the Hamiltonian \eqref{eq:Hamiltonian} can be obtained utilising the algebra of annihilation and creation operators, from which the energy levels are extracted by numerical diagonalisation. The generation of the Fock space and the numerical evaluation of the matrix elements is conceptually very easy; in Appendix \ref{Appendix:Implementation} we briefly describe our implementation. The interested reader is referred to \cite{2014JSMTE..12..010C} and \cite{Rychkov:2014eea} for further detailed discussion of the truncation approach. 

Since the Hamiltonian \eqref{eq:Hamiltonian} is not Hermitian, its spectrum can be complex in principle. Nevertheless, the spectrum is real in the case of unbroken $\op P \op T$ symmetry. However, truncation effects alter the spectrum and eigenvectors, so the truncation procedure is not guaranteed to keep the spectrum real in the $\op P \op T$ unbroken phase. Nevertheless, it turns out that the truncation procedure is safe: for reasonable cutoff values, the low-lying energy levels turn out to be real in the unbroken phase, and only become complex when $\op P \op T$ symmetry is spontaneously broken. More details on the truncation procedure are given in Appendix \ref{Appendix:Implementation}. 

\paragraph{Finite volume corrections.} When writing the Hamiltonian \eqref{eq:Hamiltonian}, we have not accounted for the mismatch between infinite volume and finite volume normal ordering. It is easy to compute the effect of normal ordering both in infinite and finite volume:
\begin{equation}
    : \varphi^2 : = \varphi^2 -Z \,,  \hspace{1 cm} : \varphi^2:_L = \varphi^2-Z_L \,,
\end{equation}
where 
\begin{equation}
    Z = \int \frac{\de k}{4 \pi} \frac{1}{\sqrt{k^2+m^2}} \,,\hspace{1 cm} Z_L = \sum_{n} \frac{1}{2 L \omega_{k_n}} \,.
\end{equation}
Although both $Z$ and $Z_L$ are ultraviolet divergent, their difference is finite \cite{Rychkov:2014eea}:
\begin{equation}
    z(L) = Z_L-Z=\frac{1}{\pi} \int_0^\infty \frac{\de x}{\sqrt{m^2 L^2+x^2}} \left(e^{\sqrt{m^2L^2+x^2}}-1\right)^{-1} \,.
\end{equation}
This implies that due to the presence of $g_2 \op V_2$ in the Hamiltonian \eqref{eq:Hamiltonian}, in finite volume there is an additional term 
\begin{equation}
    g_2 L z(L) \op 1 \,.
\end{equation}
Similarly, the free Hamiltonian contributes with a constant vacuum energy shift \cite{Rychkov:2014eea} \begin{equation}\label{eq:zeroenergy}
    E_0(L)  =  - \frac{1}{\pi L} \int_0^\infty \frac{\de x \ x^2}{\sqrt{m^2L^2+x^2}} \left(e^{\sqrt{m^2L^2+x^2}}-1\right)^{-1} \,.
\end{equation}
Similar contributions appear for the interaction terms. For the cubic potential $\op V_3$: 
\begin{equation}
    :\varphi^3: = \varphi^3-3 Z \varphi \,, \hspace{1 cm} :\varphi^3:_L = \varphi^3-3 Z_L \varphi \,,
\end{equation}
leading to
\begin{equation}
    :\varphi^3:-:\varphi^3:_L = - 3 (Z-Z_L) \varphi = 3 z(L) \varphi  \,.
\end{equation}
For the quartic term, the finite size correction is \cite{Rychkov:2014eea}:
\begin{equation}
     :\varphi^4:-:\varphi^4:_L = 6 z(L) :\varphi^2:_L+3 z(L)^2  \,.
\end{equation}
Putting everything together, we have that the Hamiltonian  \eqref{eq:Hamiltonian} must be corrected by the additive terms
\begin{equation}
     \left(E_0(L)+g_2 L z(L) +3 g_4 L z(L)^2 \right) \op 1 +3 i g_3 z(L) \op V_1+6 g_4 z(L) \op V_2 \,.
\end{equation}
Besides finite volume effects, another general issue of Hamiltonian truncation is dependence on the cutoff, which can be improved using renormalisation group methods \cite{2007PhRvL..98n7205K,2008JSMTE..03..011F,Rychkov:2014eea}; however, this turns out to be unneeded for our purposes here and so we do not go into further detail.

\paragraph{Critical points and conformal dimensions}
The vicinity of a critical point can be identified from the finite volume spectrum by searching for a point in the parameter space where the ground state meets the first excited state. However, at the eventual critical point, the two levels should approach each other for $L\rightarrow\infty$. This requires a fine-tuning of the coupling; however, due to the presence of truncation, the critical point can only be located approximately by pushing the point in which the ground state meets the first excited state to as large a volume as possible; details of this procedure are provided in \cite{Xu:2022mmw} and in \cite{Lencses:2022ira}. Then at sufficiently large values of $L$ we expect to obtain the energy spectrum of the infrared fixed point CFT, which has the form
\begin{equation}
    E_i \simeq \frac{2 \pi}{L} \left(2\Delta^\text{ir}+2n^\text{ir}-\frac{c^\text{ir}}{12}\right)+ \mathcal F L   \,,
\end{equation}
where ir stands for infrared and $\mathcal F$ is the bulk energy term. Therefore, we consider the quantities
\begin{equation}\label{eq:CiDef}
    C_i = L \frac{E_i-E_0}{4 \pi} \sim \Delta^\text{ir}-\Delta_\text{min}^\text{ir}+n^\text{ir} \,, 
\end{equation}
as a function of the volume, which must approach constant values characteristic of the infrared CFT. 

The inevitable deviation from the critical prediction can be accounted for using an effective field theory (EFT) constructed out of the least irrelevant deformations of the expected infrared CFT fixed point \cite{Xu:2022mmw,Xu:2023nke}; however, this turned out to be unnecessary in our present investigations.
 
\subsection{Testing the implementation}

\subsubsection{Quadratic perturbation}

The first non-trivial test for the Hamiltonian truncation implemented as described in the previous Section is provided by including a purely quadratic interaction: 
\begin{equation}\label{eq:HamiltonianFree}
    \op H_2 = \op H_0 + g_2 \op V_2+ \left(E_0(L)+g_2 L z(L)\right)\op 1 \,,
\end{equation}
leading to a free theory with mass
\begin{equation}
    M^2 = m^2+2 g_2 \,.
\end{equation}
where $m$ is the mass of the unperturbed theory described by $\op H_0$. We can compare the energies resulting from the Hamiltonian truncation applied to \eqref{eq:HamiltonianFree} to the energies of the free theory with mass $M$. For a full comparison, it is necessary to include a ground-state energy shift. In fact, the expected ground-state energy is 
\begin{equation}\label{eq:groundStateBogo}
    \mathcal E_0(L) = E_0(L,M) + L \Lambda   \,, \hspace{1 cm} \Lambda = \frac{1}{8 \pi} \left(M^2 \left(1-\log \frac{M^2}{m^2}\right)-m^2\right) \,,
\end{equation}
where $E_0(L,M)$ is given in equation \eqref{eq:zeroenergy}, which can be computed using a Bogoliubov transformation \cite{Rychkov:2014eea}.
\begin{figure*}[t]
\captionsetup[subfigure]{justification=centering}
\begin{subfigure}[t]{.48\textwidth}
    \includegraphics[width=0.99\textwidth]{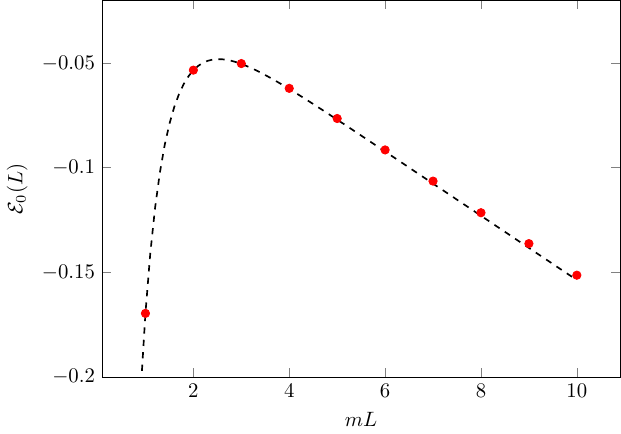}
    \caption{Ground state energy}
\end{subfigure}
\hfill
\begin{subfigure}[t]{.48\textwidth}
    \includegraphics[width=0.93\textwidth]{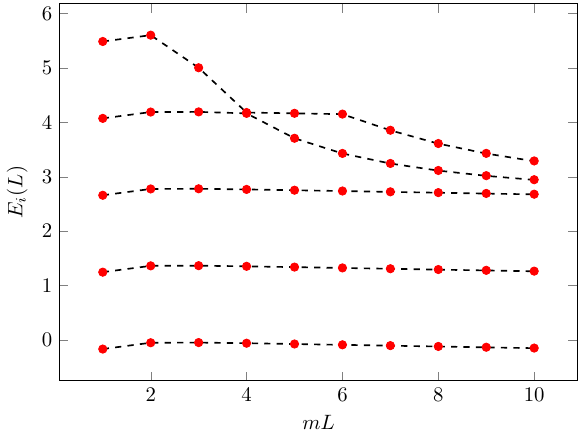}
    \caption{Excited state energies}
\end{subfigure}
  \caption{ (a) The ground state energy computed with Hamiltonian truncation (red dots) at different volumes compared to the analytical expression (dashed line). (b) Excited state energy levels from Hamiltonian truncation (red dots) compared to exact results (dashed line).}
  \label{Fig:EnergiesFreeAndDifferences}
\end{figure*}
We present results for the parameters $m = 1$, $M = \sqrt{2}$, $g = 1/2$  
in Fig. \ref{Fig:EnergiesFreeAndDifferences}, where we plot the ground state at different values of the volumes compared to the analytical prediction in \eqref{eq:groundStateBogo}, as well as the energies of a few excited states. Clearly, truncation errors increase with volume $L$. To better gauge the quality of the approximation, we include a numerical comparison $L = 1$ and $L =5$ in Table \ref{TableComparison}. 
\begin{table}
\begin{center}
\begin{tabular}{  c || c |c| c }
Volume  &Energy level& Hamiltonian Truncation & Exact  \\ \hline 
\multirow{2}{*}{$ L = 1 $} & $E_0$ &
-0.16959\ldots  &-0.16961\ldots \\
& $E_1$ &
1.24462\ldots &1.24460\ldots \\
& $E_2$ &
2.65883\ldots &2.65882\ldots \\
& $E_3$ &
4.07304\ldots &4.07303\ldots \\
& $E_4$ &
5.48726\ldots &5.48724\ldots \\
 \hline \hline 
 \multirow{2}{*}{$ L = 5 $} & $E_0$ &
-0.07647\ldots &-0.07704\ldots \\
& $E_1$ &
1.33792\ldots &1.33717\ldots \\
& $E_2$ &
2.75215\ldots &2.75139\ldots \\
& $E_3$ &
3.70771\ldots &3.70668\ldots \\
& $E_4$ &
4.16662\ldots &4.16560\ldots \\
 \hline \hline 
 
\end{tabular}
\caption{Numerical comparison of the energy levels obtained from the Hamiltonian truncation to the exact values. }
\label{TableComparison}
\end{center}
\end{table}

\subsubsection{The Ising transition and Chang duality} \label{sec:criticalCheck}
The Ginzburg--Landau Lagrangian corresponding to the Ising fixed point is given by \begin{equation}
    \mathcal L = \frac{1}{2}(\partial \varphi)^2+ \lambda \varphi^4 \,.
\end{equation}
The existence of an Ising fixed point was verified using Hamiltonian truncation in \cite{Rychkov:2014eea,Rychkov:2015vap}. We revisit this case to test our numerical implementation of the Hamiltonian truncation. 
We implemented the Hamiltonian \begin{equation}
    \op H_4 = \op H_0+ (g_2+6 g_4 z(L)) \op V_2+ g_4 \op V_4 +(E_0(L)+g_2 L z(L)+3 g_4 L z(L)^2)\op 1 \,.
\end{equation}
Since the only effect of $g_2$ is to shift the mass, we can choose $g_2 = 0$, so a single coupling parameterises the scaling region. 
It is known that a direct search for the critical point is hard to perform directly~\cite{Rychkov:2014eea,Rychkov:2015vap}. Therefore, we make use of the Chang duality of the $\varphi^4$ theory in two dimensions, which is a weak-strong duality of the theory allowing us to check the presence of the Ising fixed point for weak coupling; the price to pay is the appearance of a negative mass for the field $\varphi$. 

\paragraph{Chang duality} 
As proposed in \cite{Chang:1976ek} and numerically verified by Hamiltonian truncation in \cite{Rychkov:2015vap}, the following two Lagrangians 
\begin{equation}
    \mathcal L = :\frac{1}{2}(\partial \varphi)^2+ \frac{1}{2}m^2 \varphi^2+ g \varphi^4:_m
\label{eq:direct}\end{equation}
and
\begin{equation}
    \mathcal L' = :\frac{1}{2}(\partial \varphi)^2-\frac{1}{4}M^2 \varphi^2+g \varphi^4+\Lambda:_M
\label{eq:Chang_dual}\end{equation}
are physically equivalent provided
\begin{equation}\label{eq:Chan1}
    \frac{1}{2} m^2 + \frac{6 g}{4 \pi} \log \frac{m^2}{M^2}  = -\frac{1}{4}M^2  \,, 
\end{equation}
and
\begin{equation}\label{eq:Chang2}
    \Lambda = \frac{1}{8 \pi}\left(M^2-m^2\right)+\frac{m^2}{8 \pi}\log \frac{m^2}{M^2}+\frac{3g}{64 \pi^2}\log^2\frac{m^2}{M^2} \,.
\end{equation}
Here $:\ldots :_m$ and $:\ldots :_M$ indicate the normal ordering with respect to mass $m$ and mass $M$, respectively. The proof can be found in \cite{Chang:1976ek} and reviewed in detail in \cite{Rychkov:2015vap}, presenting a detailed numerical check of the duality itself. As a result, for some value of $g$ and a negative value of the squared mass, there is a critical point corresponding to the Ising universality class, i.e. ruled by the minimal model $\mathcal M(3,4)$. 

Since the Ising model is characterized by three Virasoro primaries $\sigma$, $\epsilon$ and $\op 1$ of conformal weights $\left (\frac{1}{16},\frac{1}{16}\right)$,  $\left (\frac{1}{2},\frac{1}{2}\right)$ and  $\left (0,0\right)$, we expect to have the following values for the $C_i$ defined in \eqref{eq:CiDef} for the first three excited energy levels:
\begin{align}
     \sigma : &\ \ \  C_1 = \frac{1}{16} \,,  \\\epsilon :& \ \ \  C_2 = \frac{1}{2} \,,
      \\L_{-1}\overline L_{-1}\sigma :& \ \ \  C_3 = \frac{17}{16}\  .
\end{align}
We found that the critical point is located at 
\begin{equation}
    g_2 \sim -0.79 \,, \hspace{1 cm} g_4 \sim 0.3 \,,  \,, 
\end{equation}
where we used units in which $m=1$. 

\begin{figure*}[t]
\captionsetup[subfigure]{justification=centering}
\centering
\begin{subfigure}[t]{.48\textwidth}
   \includegraphics[width=0.99\textwidth]{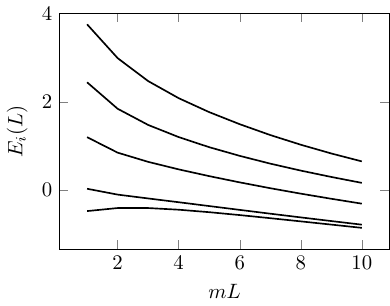}
  \caption{Energy levels close to the critical point}
	\label{fig:IsingEner}
\end{subfigure}%
\hfill
\begin{subfigure}[t]{.48\textwidth}
    \includegraphics[width=0.99\textwidth]{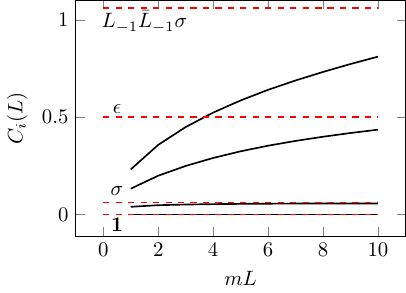}
    \caption{Determining the conformal spectrum}
	\label{fig:IsingConf}
\end{subfigure}
	\caption{ (a) Energy levels close to the critical point from Hamiltonian truncation. (b) differences between conformal dimensions defined in \eqref{eq:CiDef} near the fixed point compared to the prediction from the minimal model $\mathcal M(3,4)$, i.e. the Ising fixed point.}
  \label{Fig:IsingfixedPoint}
\end{figure*}

Fig.~\ref{Fig:IsingfixedPoint} shows the energy levels resulting from Hamiltonian truncation close to the critical point (Fig.\ref{fig:IsingEner}), while the corresponding $C_i$ are shown in Fig. \ref{fig:IsingConf}. It is clear that the Hamiltonian truncation results are compatible with the expectation from the Ising model.

To our knowledge, the Ising fixed point in the Chang dual description \eqref{eq:Chang_dual} has never been directly tested before. However, the existence of critical point in the $m^2>0$ region \eqref{eq:direct} was tested with different methods \cite{Harindranath:1988zt,Lee:2000ac,Sugihara:2004qr,Schaich:2009jk,Milsted:2013rxa,Rychkov:2014eea}. The critical values of the coupling was determined by various method (see Table 1 of \cite{Rychkov:2014eea}): the best value was obtained using RG-improved Hamiltonian method as  $g_4/m^2 \sim 2.97$. Using Chang duality, this predicts a critical point for the Chang dual description \eqref{eq:Chang_dual} at $g_4/M^2 \sim 0.26$ \cite{Rychkov:2015vap}. The value we found is $g_4/M^2 \sim 0.258$ which is in excellent agreement with the expected value. We note that a more precise analysis would be required to examine the cutoff dependence of the critical coupling. We leave this to further studies since our aim here is restricted to verifying the correctness of our implementation.

\section{Scaling region of the $i \varphi^3$ theory}\label{sec:phi3}

\subsection{General considerations}
The $i\varphi^3$ theory is the Ginzburg--Landau description of the Yang--Lee fixed point \cite{Fisher:1978pf,Cardy:1985yy}, and it corresponds to the $n =1$ case of the family \eqref{eq:proposal}.  The upper critical dimension of the potential $\varphi^3$ is $6$, and the critical exponents be studied using $\epsilon$-expansion in $d = 6-\epsilon$ \cite{Fisher:1978pf,Fei:2014yja,Fei:2014xta}. While $d=2$ (i.e. $\epsilon = 4$) is outside of the range of validity of the $\epsilon$-expansion itself, resummation techniques give a reasonable agreement with the CFT results for $d=2$ (i.e. $\epsilon = 4$) using  \cite{Kompaniets:2021hwg}. This approach also proves successful for the supersymmetric extension $\mathcal M(3,8)$ \cite{Klebanov:2022syt}.

\begin{figure*}[t]
\captionsetup[subfigure]{justification=centering}
\centering
\begin{subfigure}[t]{.48\textwidth}
   \includegraphics[width=0.99\textwidth]{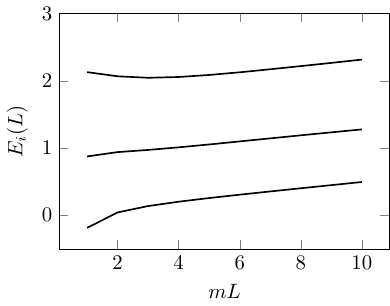}
 \caption{$\op P\op T$ symmetric phase}
	\label{fig:PTSphi3}
\end{subfigure}%
\hfill
\begin{subfigure}[t]{.48\textwidth}
    \includegraphics[width=0.99\textwidth]{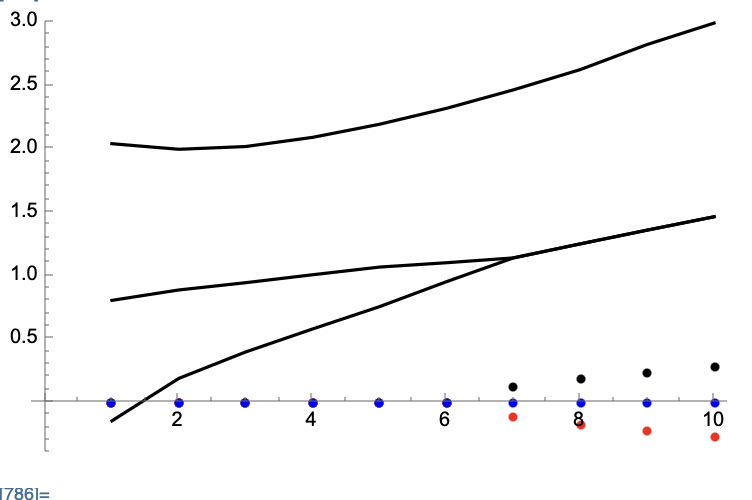}
    \caption{$\op P\op T$ symmetry breaking phase}
	\label{fig:PTBphi3}
\end{subfigure}
  \caption{(a) Spectrum of the theory described by the Hamiltonian in \eqref{eq:Hamitonianphi3} in the $\op P \op T$ symmetric phase ($g_1 \sim -0.3$, $g_2 = 0$, $g_3\sim 0.3$) the spectrum is real. (b) The spectrum of the theory described by the Hamiltonian in \eqref{eq:Hamitonianphi3} in the spontaneously broken $\op P \op T$ phase ($g_1 \sim -0.45$, $g_2 = 0$, $g_3\sim 0.3$), where complex conjugate pairs of energies appear. Solid lines correspond to the real, while dashed ones display the imaginary parts. In both cases, units are specified by setting $m=1$.}
  \label{Fig:PTphi3}
\end{figure*}

Our approach, based on Hamiltonian truncation, does not rely on the $\epsilon$-expansion and works directly in $d=2$ by implementing the theory in a finite volume $L$ with the Hamiltonian 
\begin{equation}\label{eq:Hamitonianphi3}
    \op H = \op H_0 + i \left (g_1+3 g_3 z(L)\right) \op V_1+  g_2 \op V_2 + i g_3 \op V_3+\left( E_0(L)+g_2 L z(L)\right) \op 1 \,,
\end{equation}
as discussed in the previous Section. The explicit  $\op P \op T$ symmetry or this Hamiltonian can be either realised by the eigenstates leading to a real spectrum is real, or spontaneously broken, in which case complex conjugate pairs of energies appear in the spectrum. These two cases are shown in Figs. \ref{fig:PTSphi3} and \ref{fig:PTBphi3}, respectively.

The two phases are expected to be separated by a critical point controlled by the minimal model $\mathcal M(2,5)$.
In principle, the relevant parameter space is one-dimensional. Once the quadratic coupling $g_2$  is eliminated by a suitable shift of the field $\varphi$, the scaling region is expected to be parameterised by a single relevant coupling. However, when searching for a critical point, it is best to consider both couplings $g_1$ and $g_3$ due to quantum effects that result in operator mixing. 

\subsection{The fixed point: Yang--Lee theory}
The fixed point can be found by finding the line in the space of $(g_1,g_3)$ where the ground state and the first excited state degenerate into a complex conjugate pair. This results in a line, along which further tuning must be made to push the meeting point to a large enough volume (in our case $mL> 10$) to extract the asymptotic large volume behaviour of the energies, or more precisely, the $C_i$ coefficients defined in \eqref{eq:CiDef}. 

\begin{figure*}[t]
\captionsetup[subfigure]{justification=centering}
\centering
\begin{subfigure}[t]{.48\textwidth}
   \includegraphics[width=0.99\textwidth]{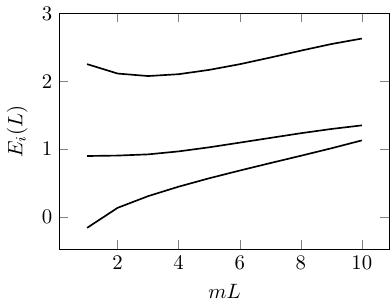}
  \caption{Lowest energy levels at the critical point}\label{fig:LeeYangEn}
\end{subfigure}%
\hfill
\begin{subfigure}[t]{.48\textwidth}
    \includegraphics[width=0.99\textwidth]{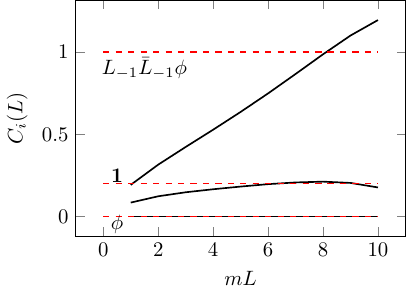}
    \caption{Determining the conformal spectrum}\label{fig:LeeYangDim}
\end{subfigure}
	\caption{(a) Spectrum of the theory described by the Hamiltonian in \eqref{eq:Hamitonianphi3} near the fixed point. The spectrum is real in the presented volume range. (b) Differences between conformal dimensions defined in \eqref{eq:CiDef} near the fixed point compared to the $\mathcal M(2,5)$ prediction.}
  \label{Fig:LeeYang}
\end{figure*}

This procedure resulted in the following estimate of the critical point: 
\begin{equation}\label{eq:phi3CP}
    g_1 \sim -0.405 \,, \hspace{1 cm} g_2 = 0 \,, \hspace{1 cm} g_3 \sim 0.4 \,, \hspace{1 cm} (\text{initial mass : }m = 1) \,.
\end{equation} 
The finite volume at this point is shown (up to volume $mL=10$) in Fig. \ref{fig:LeeYangEn}, while the $C_i$s computed from the spectrum are compared to the predictions ofthe minimal model $\mathcal M(2,5)$ In Fig. \ref{fig:LeeYangDim}. This CFT has a single Virasoro primary field (beyond the identity), whose conformal weights are $\left(-\frac{1}{5},-\frac{1}{5}\right)$, leading to the predictions: 
\begin{align}
     \op 1 :& \ \ \  C_1 = 0-\left(-\frac{1}{5}\right) = \frac{1}{5} \,, \nonumber  \\
     L_{-1}\overline L_{-1}\phi :& \ \ \  C_2 = -\frac{1}{5}+1-\left(-\frac{1}{5}\right) = 1 \label{eq:CLY}  \,  .
\end{align}

The prediction for $C_1$ agrees well with the TCSA results of Fig. (\ref{fig:LeeYangDim}). However, the result for $C_2$ seems to deviate significantly from the predicted value. This is due to an artefact of the truncation, which results in the second and the third excited states becoming a complex conjugate pair close to the Yang--Lee critical point. The two levels split into two real energy levels only at a very high cutoff. We refer the interested reader to Appendix B of \cite{Lencses:2022ira} for a description of the phenomena and also for a comparison.

As expected, the fixed point separating between the $\op P \op T$ symmetric phase and the spontaneously broken $\op P \op T$ phase is in the Yang--Lee universality class, confirming that $i \varphi^3$ is the correct Ginzburg--Landau description for the Yang--Lee fixed point. The correspondence between the primary fields of the Yang--Lee model and the fields of the Ginzburg--Landau description are given in Table \ref{Yang--Lee}.

It is possible to improve the present results by applying an effective field theory (EFT) description to match the EFT, constructed by deforming the Yang--Lee fixed point by irrelevant operators \cite{Xu:2022mmw,Xu:2023nke}. However, to have reliable results on the Wilson coefficients of the EFT, it is necessary to improve and optimise the Hamiltonian truncation implemented in this work further, which is left for future studies. 

\begin{table}[h!]
\begin{center}
\begin{tabular}{ c c c c}
\hline 
Primary & Weights & GL field & $\op P \op T$ \\ 
$\op 1$ & (0,0) &  1 & even \\
$\phi$ & (-1/5,-1/5) & $\varphi$ & odd \\
 \hline 
\end{tabular}
\caption{Primary fields in the Lee--Yang CFT $\mathcal M(2,5)$ with their conformal weights, identification in the Ginzburg--Landau description and parity under $\op P \op T$ symmetry.}
\label{Yang--Lee}
\end{center}
\end{table}

\section{Scaling region of the $\varphi^2 (i\varphi)^2$ theory}\label{sec:phi4}

\subsection{A first look at the spectrum}
\begin{figure*}[t]
\captionsetup[subfigure]{justification=centering}
\centering
\begin{subfigure}[t]{.48\textwidth}
    \includegraphics[width=0.99\textwidth]{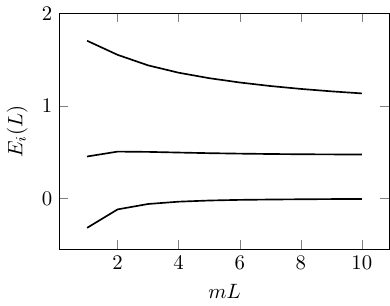}
  \caption{$\op P\op T$ symmetric phase}\label{fig:PTSphi4}
\end{subfigure}%
\hfill
\begin{subfigure}[t]{.48\textwidth}
    \includegraphics[width=0.99\textwidth]{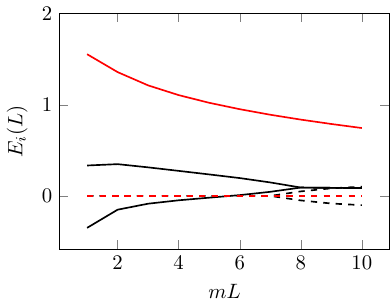}
    \caption{$\op P\op T$ symmetry breaking phase}\label{fig:PTBphi4}
\end{subfigure}
	\caption{(a) Spectrum of the theory described by the Hamiltonian in \eqref{eq:-phi4} in the $\op P \op T$ symmetric phase ($g_1 \sim -0.1$, $g_2 \sim -0.4$, $g_3= 0$, $g_4 \sim 0.3$ where the initial mass is $m = 1$): the spectrum is real. (b) Spectrum of the theory described by the Hamiltonian in \eqref{eq:Hamitonianphi3} in the spontaneously broken $\op P \op T$ phase ($g_1 \sim -0.1$, $g_2 \sim -0.52$, $g_3= 0$, $g_4 \sim 0.3$ where the initial mass is $m = 1$): in the spectrum complex conjugate pairs of energies appear, the real parts are denoted by solid and the imaginary parts with dashed lines.}
  \label{Fig:PTphi4}
\end{figure*}

Turning now to the $n=2$ case of \eqref{eq:proposal}, the corresponding finite volume Hamiltonian is 
\begin{multline}\label{eq:-phi4}
    \op H = \op H_0+ i(g_1+3 g_3 z(L) ) \op V_1+ \left(g_2+6 g_4 z(L) \right) \op V_2+ig_3 \op V_3+g_4 \op V_4+ \\ +\left(E_0(L)+g_2 L z(L) +3 g_4 L z(L)^2 \right) \op 1 \,,
\end{multline}
with $g_i \in \mathbb R$. 

Similarly to the $i\varphi^3$ case discussed in Section \ref{sec:phi3}, we expect phases with unbroken and spontaneously broken $\op P \op T$ symmetry, separated by a critical line in the universality class of the Yang--Lee model. According to the main proposal of this paper, the critical line is expected to end in the tricritical version of Yang--Lee singularity, which was found to be the minimal model $\mathcal M(2,7)$ \cite{Lencses:2023evr}. Moreover, in analogy with \cite{Lencses:2023evr}, we expect non-critical $\op P \op T$ symmetry breaking beyond the critical line's tricritical Yang--Lee endpoint. Fig. \ref{Fig:PTphi4} presents examples of the spectrum in the two phases. 

In contrast to the $i \varphi^3$ case, the $\varphi^2 (i \varphi)^2$ case has not been studied with $\epsilon$-expansion. In principle, this requires an expansion for $d= 4-\epsilon$, extending the classical work of Wilson and Fisher \cite{Wilson:1971dc} to the non-Hermitian case. However, this is a rather non-trivial task since the non-Hermitian theory requires drastically different quantization conditions for the scalar field are different, as discussed in Section \ref{sec:pt}. Additionally, getting to $\epsilon=2$ requires reaching a sufficiently high order in the expansion to make a sufficiently accurate resummation possible. As a result, the Hamiltonian truncation approach used here is much more efficient, and it is possible to establish the existence and the class of universality of the critical points as we proceed to demonstrate. 

\begin{figure}
\centering

\includegraphics[width=0.60\textwidth]{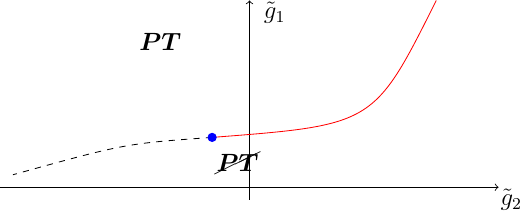}

\caption{A cartoon of the scaling region in the \textit{physical} couplings $\tilde g_1$ and $\tilde g_2$ (where the tilde refers to the couplings obtained after a proper shift the field $\varphi$ that eliminates the coupling in front $\varphi^3$ for a suitably fixed value of $g_4$). The red line is the line of critical points of the Yang--Lee type (an example of which is shown in Fig. \ref{Fig:phi4crit}), which ends in a tricritical version of the Yang--Lee singularity (shown in Fig. \ref{Fig:phi4tric}). The dashed line corresponds to non-critical $\op P \op T$ breaking (an example of which is shown in Fig. \ref{Fig:phi4tricandPT}).}

\label{fig:PhenomenologyCartoon}
\end{figure}

Generalising the $\varphi^3$ case where a single coupling parameterised the scaling region, the scaling region of $\varphi^2 (i \varphi)^2$ is spanned by two couplings. However, to reduce the problem to two independent couplings requires a shift in the field $\varphi$, which is nontrivial to parameterise since the appropriate shift depends on the couplings and the operators' mixing plays a crucial role. 
Therefore it is hard to construct explicit phase diagrams in a two-dimensional space. Nonetheless, the scaling region is expected to be analogous to Fig. 2 of \cite{Lencses:2023evr}, and we present in Fig. \ref{fig:PhenomenologyCartoon} a cartoon illustrating the expected scaling region in the space of the two independent couplings.

\subsection{The Yang--Lee critical line}

It is eventually rather easy to hit the line of Yang-Lee critical points by looking for the critical point separating the $\op P \op T$ symmetric phase from the spontaneously broken $\op P \op T$ phase. Alternatively, one can start from the case $g_4 = 0$ and  \eqref{eq:phi3CP}, then by varying $g_4$ and accordingly adjusting the other couplings, one can find the critical line. The predictions for the conformal spectrum are provided in equations \eqref{eq:CLY}.

\begin{figure*}[t]
\captionsetup[subfigure]{justification=centering}
\centering
\begin{subfigure}[t]{.48\textwidth}
   \includegraphics[width=0.99\textwidth]{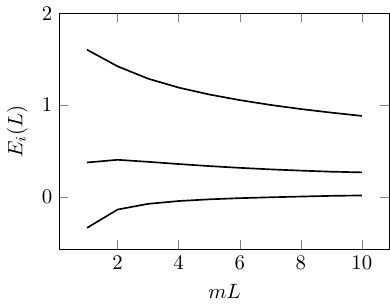}
  \caption{Energy levels near a critical fixed point}\label{fig:phi4critEn}
\end{subfigure}%
\hfill
\begin{subfigure}[t]{.48\textwidth}
    \includegraphics[width=0.99\textwidth]{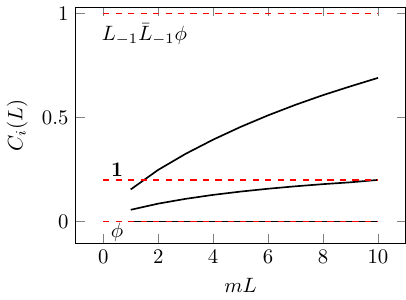}
    \caption{Determining the conformal spectrum}\label{fig:phi4critDim}
\end{subfigure}
	\caption{(a) Spectrum of the theory described by the Hamiltonian in \eqref{eq:-phi4} near a critical fixed point. (b) Differences between conformal dimensions defined in \eqref{eq:CiDef} near the same fixed point compared to the $\mathcal M(2,5)$  prediction.}
  \label{Fig:phi4crit}
\end{figure*}

The analysis of the spectrum of low-lying levels is presented in Fig. \ref{Fig:phi4crit}, where the actual point on the critical line corresponds to the couplings
\begin{equation}
    g_1 \sim -0.1 \,, \hspace{1 cm} g_2 \sim -0.48 \,, \hspace{1 cm}g_3 = 0  \  , \hspace{1 cm} g_4 \sim 0.3 \hspace{1 cm}(\text{initial mass} : m = 1)\,.
\end{equation} 
The truncation approach's numerical results clearly match the minimal model's conformal spectrum $\mathcal M(2,5)$. 

\subsection{The endpoint: tricritical version of Yang--Lee singularity}

Once a point on the critical line is found, it can be followed by tuning the couplings to find its boundary where a new critical point of a different class of universality must appear, which is expected to correspond to the minimal model $\mathcal M(2,7)$. Performing this procedure leads to the following estimate for the position of the endpoint of the critical line:
\begin{equation}
    g_1 \sim -0.115 \,, \hspace{0.5 cm} g_2 \sim -0.528 \,, \hspace{0.5 cm}g_3 = -0.24 \  , \hspace{0.5 cm} g_4 \sim 0.29 \,.
\label{eq:tricrit_couplings_estimate}
\end{equation}
where we use units in which $m=1$.

It may seem surprising that the critical point is at a positive value of the quartic coupling $g_4$. However, since the other couplings are nonzero, to determine the effective quartic coupling, one must apply the generalisation of Fisher's argument from Subsection \ref{subsec:Fisher}. Even though the couplings in the Lagrangian \eqref{eq:phi^4} and the one in \eqref{eq:tricrit_couplings_estimate} are not on the same footing since the Lagrangian couplings are bare (classical), while the couplings computed numerically are the renormalised ones, one can nevertheless insert the values \eqref{eq:tricrit_couplings_estimate} into the argument of Subsection \ref{subsec:Fisher} to estimate $\gamma_4$ in \eqref{eq:phi^4}. First, we extract $\varphi_0$ from the numerical value of the critical couplings \eqref{eq:tricrit_couplings_estimate} obtaining\footnote{One must discard possible complex solution since it was assumed that $\varphi_0$ is a real number.} $\varphi_0 \sim -0.3$. Using the relation $\gamma_4 = -15 \varphi_0^2+g_4$ gives $\gamma_4 \sim -1.15$. This is consistent with the fact that the universality class of this fixed point is different from the critical Ising (which corresponds to a positive quartic coupling), and it also confirms that the fixed point corresponds to a $\op P\op T$ invariant theory as discussed in Subsection \ref{sec:pt}, with the negative sign accounting for its non-unitarity. We also comment that the positive quartic coupling in \eqref{eq:tricrit_couplings_estimate} is important to keep the truncated Hamiltonian spectrum stable; the presence of the imaginary linear and cubic term can be interpreted as the manifestation of the nontrivial $\op P\op T$-symmetric quantisation condition known from the quantum mechanical studies.

\begin{figure*}[t]
\captionsetup[subfigure]{justification=centering}
\centering
\begin{subfigure}[t]{.48\textwidth}
   \includegraphics[width=0.99\textwidth]{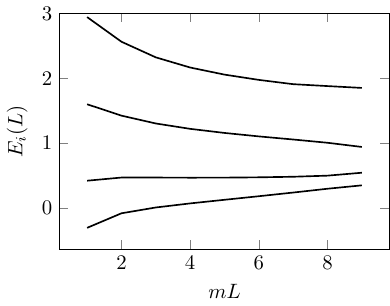}
  \caption{Energy levels near the tricritical fixed point}\label{fig:phi4tricEn}
\end{subfigure}%
\hfill
\begin{subfigure}[t]{.48\textwidth}
    \includegraphics[width=0.99\textwidth]{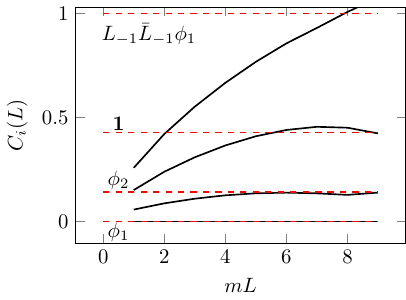}
    \caption{Determining the conformal spectrum}\label{fig:phi4tricDim}
\end{subfigure}
	\caption{(a) Spectrum of the theory described by the Hamiltonian in \eqref{eq:-phi4} near the non-unitary tricritical fixed point. (b) Differences between conformal dimensions defined in \eqref{eq:CiDef} near the non-unitary tricritical fixed point compared to the $\mathcal M(2,7)$ prediction.}
  \label{Fig:phi4tric}
\end{figure*}

To identify the fixed point, recall that the minimal model $\mathcal M (2,7)$ has three primary fields: the identity $\op 1$ of weights $(0,0)$, and two nontrivial fields $\phi_1$ and $\phi_2$ whose conformal weights are $\left(-\frac{2}{7},-\frac{2}{7}\right)$ and $\left(-\frac{3}{7},-\frac{3}{7}\right)$. Therefore we expect to find 
\begin{align}
     \phi_1 :& \ \ \  C_1 = -\frac{2}{7}-\left(-\frac{3}{7}\right) = \frac{1}{7} \,, \label{eq:C127}  \\\op 1 :& \ \ \  C_2 = 0-\left(-\frac{3}{7}\right) = \frac{3}{7} \label{eq:C227}  \  ,  \\L_{-1}\overline L_{-1} \phi_2 :& \ \ \  C_3 = -\frac{3}{7}+1-\left(-\frac{3}{7}\right) = 1 \,. \label{eq:C327}
\end{align}

Those predictions can be compared with numerical results obtained from the Hamiltonian truncation at the point \eqref{eq:tricrit_couplings_estimate}. In Fig. \ref{Fig:phi4tric}, we compare these predictions with the numerical values resulting from the Hamiltonian truncation. The resulting match confirms the presence of a critical point in the universality class of the minimal model $\mathcal M(2,7)$. 

The identification between the primary fields of $\mathcal M(2,7)$ and the Ginzburg--Landau fields can be fixed using their transformation properties under the $\op P \op T$ symmetry and is presented in Table \ref{M27}.

\begin{table}[h!]
\begin{center}
\begin{tabular}{ c c c c}
\hline 
Primary & Weights & GL field & $\op P \op T$ \\ 
$\op 1$ & (0,0) &  1 & even \\
$\phi_1$ & (-2/7,-2/7) & $\varphi$ & odd \\
$\phi_2$ & (-3/7,-3/7) & $:\varphi^2:$ & even\\
 \hline 
\end{tabular}
\caption{Primary fields in the Lee Yang model $\mathcal M(2,7)$, listing their conformal weights, their identification in the Ginzburg--Landau description and their transformation property under $\op P \op T$ symmetry.}
\label{M27}
\end{center}
\end{table}

\subsection{Non-critical $PT$ breaking}

As pointed out in \cite{Lencses:2023evr}, the absence of an order parameter for the $\op P \op T$ symmetry breaking opens the possibility for a non-critical symmetry breaking \cite{Lencses:2023evr}. The possible options for the phenomenology of $\op P \op T$ symmetry breaking are the following: 
\begin{itemize}
    \item[$\star$] The ground state meets the first excited state, forming a complex conjugate pair, which is just the critical $\op P \op T$ breaking scenario.
    \item[$\star$] The second possibility is that the ground state simultaneously meets the first and second excited states, which happens at the tricritical point. Note that, in principle, it is possible to have more lines meeting simultaneously with the ground state, which corresponds to higher multicritical points, but the $\varphi^2 (i \varphi)^2$ model does not have enough tunable parameters to tune to reach a tetracritical point\footnote{To reach such a tetracritical point it is necessary to add a term of the form $\varphi^2 (i \varphi)^3$ in accordance with the proposal \eqref{eq:proposal}.}.
    \item[$\star$] The last possibility is that the first excited state meets the second excited state forming a complex conjugate pair before meeting the ground state. Since $\op P \op T$ is spontaneously broken without closing the gap, this corresponds to a non-critical transition.
\end{itemize}

\begin{figure*}[t]
\centering
    \includegraphics[width=0.55\textwidth]{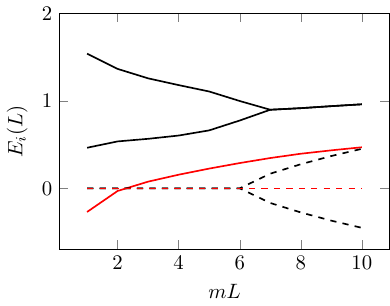}
    
  \caption{ An example of a point in the parameter space ($g_1 \sim -0.13$, $g_2 \sim -0.54$, $g_3= -0.25$, $g_4 \sim 0.2$ where the initial mass is $m = 1$) in which $\op P \op T$ symmetry is broken non-critically.}
  \label{Fig:phi4tricandPT}
\end{figure*}

Indeed, continuing beyond the endpoint of the critical line, we find a non-critical transition separating the $\op P \op T$ symmetric and symmetry-breaking regimes. An example of such a transition point in the scaling region defined by the Hamiltonian \eqref{eq:-phi4} is shown in Fig. \ref{Fig:phi4tricandPT}. It is an interesting open problem to understand this phenomenon that recently appeared in other models as well \cite{Delouche:2023wsl,Ambrosino:2023dik}.

\section{Conclusions and outlook}\label{sec:discussion}

In this work, we proposed a Ginzburg--Landau description for the non-unitary sequence of minimal models $\mathcal M(2,2n+3)$. The corresponding GL Lagrangians are the field-theoretic generalisations of the $\op P \op T$ symmetric quantum mechanical Hamiltonians proposed originally in \cite{Bender:1998ke}, which have real spectra despite their non-Hermitian nature.

According to our proposal, the Ginzburg--Landau description for the minimal model $\mathcal M(2,2n+3)$ is a single-scalar boson Lagrangian where potential has the leading term $\varphi^2(i \varphi)^{n}$. We supported our conjecture by adopting Fisher's construction for the Yang--Lee model \cite{Fisher:1978pf} and using information from integrable off-critical deformations, plus known facts related to $\op P \op T$ symmetry. 

We also performed a numerical analysis based on Hamiltonian truncation for the simplest cases $n=1$ and $n=2$, which correspond to the minimal models $\mathcal M(2,5)$ and $\mathcal M(2,7)$. After testing the implementation, which included a non-trivial identification of the Ising fixed point in the Chang dual channel, we located the critical points with the appropriate universality classes in the $i \varphi^3$ and $\varphi^2(i \varphi)^2$ theories, confirming their nature by a numerical analysis of their spectra.

Note that $\op P \op T$ symmetry was central to our discussion. In fact, the critical points we found always separate a $\op P \op T$ symmetric phase from a spontaneously broken $\op P \op T$ phase. Furthermore, we provided numerical evidence for non-critical $\op P \op T$ breaking in the scaling region of $\varphi^2 (i \varphi)^2$ theory.  Interestingly the same type of phenomenology emerges in other two-dimensional models \cite{Lencses:2022ira,Delouche:2023wsl,Ambrosino:2023dik}, which makes it interesting to understand the underlying physics in more detail.

In fact, $\op P \op T$ symmetric models have been proposed to describe actual physical phenomena \cite{Soley:2022dnl}, and experimental measurements of the Yang--Lee zeros were also proposed~\cite{PhysRevLett.109.185701,Shen:2023tst,Li:2022mjv,Matsumoto:2022}. 

As we discussed, the theories described by Lagrangians of the form of equation \eqref{eq:proposal} require specific quantisation conditions \cite{Bender:1998ke}, which restrict the usefulness of mean-field approaches and $\epsilon$-expansions, in contrast to the unitary case. 
A notable exception is the $i \varphi^3$ case, where the quantisation conditions of the theory coincide with the usual ones, and indeed our results are in perfect agreement with $\epsilon$-expansions. On the contrary, in the $\varphi^2(i \varphi)^2$ case, where the quantisation conditions are expected to differ from the usual, we can still establish the existence of critical points to which the usual $\epsilon$-expansion is completely blind.

To avoid the problem of the quantisation condition, an expansion of the potential $\varphi^2 (i \varphi)^n$ in $n$ was proposed in \cite{Bender:2006wt,Bender:2020gbh,Bender:2021fxa}. It would be interesting to understand if there is a way to modify the usual procedure of the $\epsilon$-expansion to recover the critical behaviour found here.

The Yang--Lee universality class was also extended to higher dimensions, and similar attempts were made in the case of the universality class of the minimal model $\mathcal M(3,8)$ \cite{Nakayama:2021zcr,Klebanov:2022syt}.  A natural question is whether the multi-critical Yang--Lee universality classes, described by the minimal models $\mathcal M(2,2n+3)$ (for $n>1$) in two dimensions, can be extended to higher dimensions.

Another interesting direction is to generalise the numerical approach of this paper to the case of multi-field Hamiltonians, which is in principle, possible. This is interesting in the light of related results \cite{Klebanov:2022syt,Nakayama:2022svf}, and may also lead to new Ginzburg--Landau theories for other non-unitary minimal models.

An open question is to find a proper generalisation of Zamolodchikov's OPE-based argument for the Ginzburg--Landau descriptions of unitary minimal models \cite{Zamolodchikov:1986db} to the case of non-unitary models, which is not clear at this time, despite an attempt given in Appendix A of \cite{Lencses:2022ira}.

\begin{acknowledgments}
It is a pleasure to thank D. Sz\'asz-Schagrin for very useful discussions. AM  have benefited from the German Research Foundation DFG under Germany’s Excellence Strategy – EXC 2121 Quantum Universe – 390833306. GM acknowledges the grants PNRR MUR Project PE0000023- NQSTI and PRO3 Quantum Pathfinder.  GT was partially supported by the Ministry of Culture and Innovation and the National Research, Development and Innovation Office (NKFIH) through the OTKA Grant K 138606 and also under Grant Nr. TKP2021-NVA-02. This collaboration was partly supported by the CNR/MTA Italy-Hungary 2023-2025 Joint Project “Effects of strong correlations in interacting many-body systems and quantum circuits”.
ML was partially supported by the Ministry of Culture and Innovation and the National Research, Development and Innovation Office (NKFIH) through the OTKA Grant K 134946 and the New National Excellence Program under the Grant Nr. ÚNKP-23-5-BME-456. ML was also supported by the Bolyai János Research Scholarship of the Hungarian Academy of Sciences.
\end{acknowledgments}

\appendix
\section{Explicit interaction terms for Hamiltonian truncation}\label{sec:appendixTerms}
To implement the Hamiltonian truncation, it is necessary to write explicit expressions for the terms in the potential in terms of the annihilation and creation operators of the field $\varphi$, given in equation \eqref{eq:phimodes}. Since we are interested in powers of the field $\varphi$ up to $\varphi^4$, we write below explicitly the terms that define the Hamiltonian in \eqref{eq:Hamiltonian}.
\begin{equation}
   \op  V_1 = \int_0^L \varphi \ \de x = \sum_{k} \sqrt{\frac{L}{2 \omega_k}} \left(\op a_k+\op a_{k}^\dagger\right) \,,
\end{equation}
\begin{equation}
    \op V_2 =  \int_0^L :\varphi^2:_L \ \de x  = \sum_k \frac{1}{2 \omega_k} \left(\op a_k\op a_{-k}+ \op a_k^\dagger \op a_{-k}^\dagger+2 \op a_{k}^\dagger \op a_{k}\right)\,,
\end{equation}
\begin{multline}
    \op V_3 = \int_0^L :\varphi^3:_L \ \de x = \sum_{k_1,k_2}\frac{\op a_{k_1}\op a_{k_2}\op a_{-k_1-k_2}+\op a_{k_1}^\dagger\op a_{k_2}^\dagger\op a_{-k_1-k_2}^\dagger}{2\sqrt{2L}\sqrt{\omega_{k_1}\omega_{k_2}\omega_{-k_1-k_2}}}\\ \ \  +\sum_{k_1,k_2}\frac{\op a_{k_1+k_2}^\dagger\op a_{k_1}\op a_{k_1}+\op a_{k_1}^\dagger\op a_{k_2}^\dagger\op a_{k_1+k_2}}{\sqrt{2L}\sqrt{\omega_{k_1}\omega_{k2}\omega_{k_1+k_2}}}\,,
\end{multline}
\begin{multline}
    \op V_4 = \int_0^L :\varphi^4:_L \ \de x = \sum_{k_1,k_2,k_3}\frac{\op a_{k_1}\op a_{k_2}\op a_{k_3}\op a_{-k_1-k_2-k_3}+\op a_{k_1}^\dagger\op a_{k_2}^\dagger\op a_{k_3}^\dagger\op a_{-k_1-k_2-k_3}^\dagger}{4 L\sqrt{\omega_{k_1}\omega_{k_2}\omega_{k_3}\omega_{-k_1-k_2-k_3}}} \\ \ \  +\sum_{k_1,k_2,k_3}\frac{\op a_{k_1+k_2+k_3}^\dagger\op a_{k_1}\op a_{k_2}\op a_{k_3}+\op a_{k_1}^\dagger\op a_{k_2}^\dagger\op a_{k_3}^\dagger\op a_{k_1+k_2+k_3}}{L\sqrt{\omega_{k_1}\omega_{k_2}\omega_{k_3}\omega_{k_1+k_2+k_3}}}+\\ +3\sum_{k_1,k_2,k_3}\frac{\op a_{k_1}^\dagger \op a_{k_2}^\dagger \op a_{k_3}\op a_{k_1+k_2-k_3}}{2 L\sqrt{\omega_{k_1}\omega_{k_2}\omega_{k_3}\omega_{k_1+k_2-k_3}}} \,.
\end{multline}
\section{Implementation of the Hamiltonian truncation}\label{Appendix:Implementation}
Here, we give some details on the implementation with suggestions for its optimisation. 
\paragraph{Basis generation:} The first step is to generate the basis of the truncated Fock space of the free massive theory. We first generate all the possible single-particle states below the chosen energy cutoff and then construct all the combinations of those single-particle states with only positive momenta, which satisfy the energy cutoff, providing a basis for right-movers; left movers can be obtained by flipping all particle momenta negative. Then we construct a zero-momentum subspace by taking all possible combinations of the right-mover states below the imposed energy cutoff. Finally, the full basis is constructed by adding zero momentum particles so that the resulting states stay below the energy cutoff.
\paragraph{Matrix element computation:} Due to the creation/annihilation operators' action, the interaction terms $V_n$ matrices are very sparse. The generation of the matrix elements can be optimised by running over all the states in a single loop and determining the list of vectors in the truncated basis produced from each basis vector by the action of $V_n$. Then we compute the matrix element with the initial basis vector for each such vector, thereby obtaining the matrix elements in a form suitable for sparse matrix storage.
\paragraph{Hamiltonian construction:} note that the generation of the truncated basis and the matrix elements of the interaction operators $V_n$ must be run only once for each volume value since these data are independent of the coupling. Therefore, the eventual Hamiltonian can be computed by a linear combination of these matrices weighted with the desired values of the couplings. 

\paragraph{Non-hermiticity, $P T$ symmetry and stability of truncation}: For the $\op P \op T$ symmetric non-Hermitian Hamiltonians considered in this paper, the reality of the spectrum is guaranteed in $\op P \op T$ symmetric phase. However, Hamiltonian truncation generally spoils the reality of the spectrum. Indeed, preliminary studies of the quantum mechanical Hamiltonian \begin{equation}\label{ham:x3}
    \op H = p^2+i x^3
\end{equation}
using a simple Hamiltonian truncation explained in chapter 25 of \cite{Mussardo:2020rxh}, show that while the first few eigenvalues are real, to extend the reality of the spectrum for higher eigenstates requires a relatively high energy cutoff and, therefore, a large number of states. To show this, we implemented the Hamiltonian truncation keeping $100$ states and $5000$ states (the latter is the order of magnitude of the number of states used in the field theoretical counterpart implemented in this paper), and we only show the first four energy levels (see Tab. \ref{tab:qmtruncation}). Note that the third and fourth states are complex when the number of states is $100$ but become real when the cutoff increases.
\begin{table}
\begin{center}
\begin{tabular}{  c || c |c }
\hline \hline 
\multirow{2}{*}{100 states}
& $ E_0$ &
$1.15627$  \\
& $ E_1$ &
$4.10923$  \\
& $ E_2$ &
$6.08481 +2409.48 i$  \\
& $ E_3$ &
$6.08481 -2409.48 i$  \\
 \hline \hline 
 \multirow{2}{*}{5000 states} 
& $ E_0$ &
$1.15627$  \\
& $ E_1$ &
$4.10923$  \\
& $ E_2$ &
$7.56227$\\
& $ E_3$ &
$11.3144 +10^{-10} i$  \\
 \hline \hline 
\end{tabular}
\caption{Lower energy levels of the quantum mechanical model described by the Hamiltonian \eqref{ham:x3} for different cutoffs. The imaginary part is omitted when it is of order $\mathrm O(10^{-11})$.}
\label{tab:qmtruncation}
\end{center}
\end{table}

Fortunately, it turns out that the field-theoretic version does not suffer additional problems. We tested that the reality of the spectrum is stable under the truncation by explicitly computing the imaginary part of the energies. We give the relative imaginary parts of the energies, i.e.
\begin{equation}
    \delta E_i = \frac{\operatorname{Im}E_i}{|E_i|}\,,
\end{equation}
in Table \ref{tab:im} for the choice of the couplings of Fig. \ref{fig:PTSphi3} and Fig.  \ref{fig:PTSphi4}, where the exact spectrum is expected to be real. Since $\delta E_i$ depends on the volume, we give its maximum value for the volume range considered. It is clear that the reality of the spectrum holds with a very high numerical precision.
\begin{table}
\begin{center}
\begin{tabular}{  c || c |c }
\hline \hline 
\multirow{2}{*}{$ \varphi^3 $}
& $\delta E_0$ &
$3 \cdot 10^{-11}$  \\
& $\delta E_1$ &
$1 \cdot 10^{-10}$  \\
& $\delta E_2$ &
$-2 \cdot 10^{-10}$  \\
 \hline \hline 
 \multirow{2}{*}{$ \varphi^2(i \varphi)^2 $} 
& $\delta E_0$ &
$6 \cdot 10^{-14}$  \\
& $\delta E_1$ &
$9 \cdot 10^{-15}$  \\
& $\delta E_2$ &
$1 \cdot 10^{-14}$  \\
 \hline \hline 
\end{tabular}
\caption{Imaginary part of the energy divided by its absolute value for $\op P \op T$ unbroken phases of the $i\varphi^3$ and $\varphi^2 (i \varphi)^2$ GL models. The choice of the couplings is the same as in Figs. \ref{fig:PTSphi3} and \ref{fig:PTSphi4}, respectively.}
\label{tab:im}
\end{center}
\end{table}
\newpage
	 \bibliographystyle{JHEP}
	 \bibliography{GL}
\end{document}